\documentclass{emulateapj}
\usepackage{amssymb}
\usepackage{graphicx}
\usepackage{color}

\def\gsim{\lower 2pt \hbox{$\, \buildrel {\scriptstyle >}\over
{\scriptstyle \sim}\,$}}
\def\lsim{\lower 2pt \hbox{$\, \buildrel {\scriptstyle <}\over
{\scriptstyle \sim}\,$}}
\def\hea{He{$\alpha$}}
\def\lya{Ly{$\alpha$}}

\def\oviii{O{\small VIII}}
\def\ovii{O{\small VII}}

\def\neix{Ne{\small IX}}

\def\nex{Ne{\small X}}

\def\fexvii{Fe{\small XVII}}

\def\etal{et~al.}

\def\cmsq{\hbox{cm$^{-2}$}}
\def\chandra{{\sl Chandra}}
\def\kmps{\hbox{km $\rm{s^{-1}}$}}

\newcommand{\ssst}{\scriptscriptstyle}
\newcommand{\E}[1]{\times 10^{#1}}

  \newcommand{\ps}{\,{\rm s}^{-1}}
    
\newcommand{\cm}{\,{\rm cm}}    \newcommand{\km}{\,{\rm km}}

\newcommand{\K}{\,{\rm K}}      
\newcommand{\keV}{\,{\rm keV}}

\newcommand{\no}{n_{\ssst 0}}

\newcommand{\sigth}{\sigma_{\rm th}}    \newcommand{\sigtu}{\sigma_{\rm tu}}
    \newcommand{\nH}{n_{\ssst\rm H}}

\newcommand{\rc}{r_{\rm c}} \newcommand{\rcut}{r_{\rm cut}}

\newcommand{\ka}{K$\alpha$}

\newcommand{\XMM}{{\sl XMM-Newton}}

\shorttitle{Resonant Scattering in Hot ISM}

\begin{document}

\title{Resonant scattering effect on the soft X-ray line emission from the hot interstellar medium: I. galactic bulges}

\author{Yang Chen\altaffilmark{1,2}, Q.\ Daniel Wang\altaffilmark{3}, Gao-Yuan Zhang\altaffilmark{1}, Shuinai Zhang\altaffilmark{3,4}, and Li Ji\altaffilmark{4} }
\altaffiltext{1}{Department of Astronomy, Nanjing University, 163 Xianlin Avenue, Nanjing 210023,
       P.R.China}
\altaffiltext{2}{Key Laboratory of Modern Astronomy and Astrophysics, Nanjing University, Ministry of Education,
       P.R.China}
\altaffiltext{3}{Department of Astronomy, B619E-LGRT,
      University of Massachusetts, Amherst, MA01003}
\altaffiltext{4}{Purple Mountain Observatory, 8 Yuanhua Road, Nanjing 210034, P.R.China}

\begin{abstract}
Diffuse soft X-ray line emission is commonly used to trace the thermal
and chemical properties of the hot interstellar medium, as well as its
content, in \mbox{nearby} galaxies. Although
resonant line scattering
complicates the interpretation of the emission, it
also offers an opportunity to measure the kinematics of the medium.
We have implemented a direct Monte Carlo simulation scheme that enables us
to account for resonant scattering effect in the medium, in principle,
with arbitrary spatial, thermal, chemical, and kinematic distributions.
Here we apply this scheme via dimensionless calculation to
an isothermal, chemically uniform, and spherically symmetric medium
with a radial density distribution characterized by a $\beta$-model.
This application
simultaneously account for both optical depth-dependent spatial distortion and
intensity change of the resonant line emission due to the scattering,
consistent with previous calculations. We further
apply the modeling scheme to the \ovii\ and \oviii\
emission line complex observed in the  \XMM\ RGS spectrum of the M31 bulge.
This modeling, though with various limitations due to its simplicity, shows
that the resonant scattering could indeed account for much of the
spatial distortion of the emission, as well as the relative strengths of the lines, especially
the large forbidden to resonant line ratio of the \ovii\ \hea\ triplet.
We estimate the isotropic turbulence Mach number
of the medium in M31 as $\sim0.17$
 for the first time and the line-emitting gas temperature
as $\sim2.3\times10^6$\,K.
We conclude that the resonant scattering may in general play an
important role in shaping the soft X-ray spectra of diffuse hot gas
in normal galaxies.

\end{abstract}

\keywords{X-rays: galaxies ---
          galaxies: bulges ---
          galaxies: individual (M31) ---
          X-rays: ISM ---
          ISM: lines and bands ---
          radiative transfer
}

\section{Introduction}
The hot interstellar medium (ISM) represents a
key component of the galactic ecosystem in disk galaxies similar to our own. In particular, the medium bears
direct imprints of stellar feedbacks, both energetical and chemical, which
are an essential ingredient that is yet to be understood in the theory
of galaxy formation and evolution.
With a characteristic temperature of $\gtrsim 10^6\K$, the medium
can be well traced by soft X-ray emission, which is
dominated by various lines from K- or L-shell transitions of
heavy elements such as O, Ne, and Fe ions (e.g., \ovii, \oviii, \neix,
and \fexvii). The emission is commonly assumed to be optically thin
in spectral modeling. However, this assumption may not hold
for certain resonant lines,
which depend on their oscillation strengths,
the thermal broadening and turbulent velocity dispersion and density
of the hot gas, and the abundances and ionization states
of the elements.


A photon in a resonant line can be absorbed and then
re-emitted by a suitable ion in different directions.
In such a resonant ``scattering'' (RS) event, the energy of
this photon is shifted, mainly due to the turbulent and thermal
random motions of the ion. Statistically and cumulatively, this shift
tends to be away from the line center, where the
optical depth is the highest, and increases the chance for the photon to
escape from the medium. The resultant line broadening is sometimes called
Zanstra effect (see, e.g., Zanstra 1949; Field 1959),
the direct measurement of which, however, demands very high
spectral resolution observations. 
With the existing X-ray instruments, we can detect the spatial distortion
of the surface intensity distribution of the line emission due to the RS effect.
The degree of this distortion
is optical depth-dependent; the observed line photons primarily escape
from outer regions of the medium, which are optically thin to the line.
The relative intensities of various emission lines are altered in
a spectrum extracted from a fixed field covering only part of the medium.
In addition, the RS
increases the accumulated propagation path length of a photon and hence
its probability for being absorbed photoelectrically by cool gas
if it coexists with the hot medium. Accounting for
these effects is thus essential
for the measurements of
the thermal and chemical properties of the medium, which are sensitive
to the line intensities. Incidentally, the line optical depth dependence of the
RS effects can provide us with a powerful tool to constrain the
kinematic state of the medium.

The RS effect in X-rays has been studied mostly in diffuse hot plasma of
massive elliptical galaxies and clusters of galaxies
(e.g., Xu et al.\ 2002; Sazonov, Churazov, \& Sunyaev 2002; Churazov et al.\ 2004;
Molnar, Birkinshaw, \& Mushotzky 2006; Werner et al.\ 2009;
Zhuravleva et al.\ 2011; Ogorzalek et al.\ 2017; Hitomi Collaboration et al.\ 2018).
The spectra of diffuse
X-ray emission from such systems typically show distinct
iron emission lines from K-shell and L-shell transitions.
The relative intensities
of the emission lines of vastly different opacities, arising from different
transitions of the same species (e.g., Fe XVII 15 and 17 \AA\ lines),
have effectively been used to infer the turbulent Mach numbers in the
plasma, which are up to a few tenths.
The earliest theoretical work by Gilfanov, Syunyaev, \& Churazov (1987) is based on an
iteration method to estimate the radiative transfer equation, 
 assuming an isothermal spherical $\beta$-model gas
distribution, and neglecting the photo-electric absorption by cool gas,
which might be intermixed with the hot medium.
 Later works typically use the Monte Carlo (MC)
radiative transfer simulations of the resonant scattering effect on lines
typical for the $\ga10^7\K$ gas
(see, e.g., the other references mentioned in this paragraph).

In disk galaxies, the RS can also be important, although its effects may not
be separated easily from other physical processes in the hot ISM.
The characteristic
temperature of the hot ISM is $\sim 10^{6.3}$~K, as estimated from existing
X-ray absorption and emission observations (e.g., Wang 2010).
At such a temperature, the medium is expected to produce strong
line emission from K-shell transitions in H-like and He-like ions,
such as \oviii, \ovii, \nex, and \neix, as well as those Fe L-shell
ones (e.g.,  Wang 2010; Porquet \etal\ 2010; Foster \etal\ 2012).
Although the collisional ionization equilibrium (CIE) is often {\sl assumed}
for the hot ISM, complications can arise from non-CIE processes
such as the photo-ionization by recent or ongoing AGN activity (e.g., Segers \etal\ 2017)
and the charge exchange (CX) between ions and neutral atoms
at the interface with cool gas (Ranalli \etal\ 2008; Liu \etal\ 2010, 2011; Zhang \etal\ 2014), as well as the over-heating
(e.g., due to recent shock-heating) or over-cooling (e.g.,
fast adiabatic expansion; e.g., Breitschwerdt \& Schmutzler 1999). 
These processes, which could be common in the dynamic (or violent) hot ISM of
star-forming disk galaxies, are important to study.


Despite these potential complications, observational signs of significant RS effects are present for nearby disk galaxies, including our own.  X-ray
{\sl absorption} line spectroscopy of bright
background point-like sources reveals that the hot ISM  in our Galaxy
is optically thick to the resonant line of the \ovii\
\hea\ triplet (e.g., Wang et al. 2005). The RS by the foreground
hot ISM is also used to explain the emission line intensity ratios
observed in the {\sl Suzaku} spectrum of the North Polar Spur (Gu \etal\ 2016).
The \XMM\ RGS study of the bulge of M31 shows that the forbidden to
resonant line ratio of the \ovii\ \hea\ triplet is much greater than 1, as
would be expected from a hot CIE plasma (Liu \etal\ 2010; Zhang \etal\ 2018a, in prep). There is no significant massive star
formation or ongoing AGN activity in the bulge, which contains
only a small amount of cool gas, mainly confined in the inner 200~pc.
In such a ``quiescent'' environment, a significant CX is not
expected\footnote{
Liu et al.\ (2010) qualitatively considered CX as an interpretation of the enhanced \ovii\ K$\alpha$ forbidden line. This interpretation seems to be consistent with the similarity between the radial intensity distributions of the line and the H$\alpha$ emission in the inner bulge of M31. But, quantitatively, the interpretation might have difficulties, because the interface between the cool and hot gases in the M31 inner bulge, where H$\alpha$ emission is confined in a flattened spiral structure (Dong \etal\ 2016), is likely to be substantially small and weakly turbulent (Zhang et al.\ 2018a in prep).}.
Can the RS play a role? Based on existing studies
(Li \etal\ 2009; Liu \etal\ 2010), we find
that the inner bulge region of M31
could indeed be substantially
optically thick in strong resonant lines of \ovii, \oviii,
and Fe{\small XVII}, for example,
depending largely on how small the turbulent velocity of the
diffuse hot ISM is. This dependence can in turn be used to estimate
the velocity, which is crucially needed for the study of the energetics and
dynamics of the hot ISM.

The RS effect on the X-ray spectroscopic properties of the
hot ISM in disk galaxies has not been well explored yet
since the modeling is somewhat challenging.
The optical depth is
too large ($\ga$1) for the RS to be treated as a simple correction, but
is not large enough to be considered as a diffusion problem.
The mean free path of line photons could be comparable to, or even
larger than, the characteristic length scale of the hot ISM.
Thus MC simulations should be
an optimal approach to quantify the RS effects, allowing for
complications due to various spatial, thermal, chemical,
and kinematical properties of the hot ISM.
Indeed, this MC approach has been
used successfully for similar radiative transfer problems
(e.g., 
Ahn, Lee, \& Lee 2000; Zheng \& Miralda-Escud{\'e} 2002;
Cantalupo et al.\ 2005; Dijkstra, Haiman, \& Spaans 2006a,b)
and especially in dealing with the RS effect on the X-ray emission of hot intracluster gas
(e.g., Sazonov et al.\ 2002; Churazov et al.\ 2004;
Molnar et al.\ 2006; Werner et al.\ 2009; Zhuravleva et al.\ 2010; Zhuravleva et al.\ 2011;
de Plaa et al.\ 2012; Zhuravleva et al.\ 2013; Ogorzalek et al.\ 2017)
over a decade or two.

In this paper we present our calculation scheme of the MC simulation for the
radiative transfer of the resonant lines in hot gas, the spatial distribution
of which is characterized by the $\beta$-model.
We largely follow the formulism described in Dijkstra et al.\
(2006a) and Zheng \& Miralda-Escud{\'e} (2002) for the RS
of hydrogen Ly$\alpha$ photons and adapt it to the
RS of the soft X-ray line emission from the hot ISM.
We describe the scheme in some algorithmic details,
which incorporates a finite boundary, as well as the transformation between
the ion's rest frame and the laboratory frame
and the ion velocity-dependent probability and recoil effect of the RS.
Our focus here is on an
application of the scheme to a simple case, in which the hot ISM is assumed
to be isothermal, chemically uniform, and spherically symmetric.
This application allows us to directly compare some of
our results  with those obtained by the iterative method (Gilfanov \etal\ 1987)
and apply them to the analysis of the RGS data on the inner bulge of M31.
In subsequent papers, we will explore more sophisticated cases (e.g.\ Zhang
et al.\ 2018b in prep, Paper II), including the RS in non-spherical
asymmetric gaseous spheroids or disks, the effect due to
photo-electric absorption, and the inclination angle dependence, as well as
the different viewing perspectives, internal or external to the medium.

The rest of the present paper is organized as follows: In \S~2, we describe
the modelling and the MC algorithm for the RS in the hot ISM of galactic bulges.
We present our test simulations in \S~3 and an application to the X-ray spectrum
of the M31 bulge in \S~4. We discuss the results in \S~5 and summarize this work in \S~6.

\section{Resonant scattering Modeling} \label{sec:modeling}

\subsection{Fundamentals}

\begin{center}
\begin{deluxetable*}{lcccc|cc}
\tablewidth{0pt}
\tablecaption{The Line properties and optical depths fitted to the data}
\tablehead{
Ion $(\lambda_{\rm rest})$  & transition  & $f_{lu}$ & $\Gamma$ & $w_{\rm I}:w_{\rm D}$ & \multicolumn{2}{c}{$\tau^{\dagger}$} \\
$\qquad\ [\textrm{\AA}]$    &      &     &  s$^{-1}$ & &  }
\startdata
\oviii~Ly$\alpha$ (18.967)  & 2p\,$^2$P$_{3/2}$--1s\,$^2$S$_{1/2}$ & 0.28  & $2.566\times10^{12}$ & 1/2:1/2 & 4.80 & 4.19  \\
\oviii~Ly$\alpha$ (18.973)  & 2p\,$^2$P$_{1/2}$--1s\,$^2$S$_{1/2}$ & 0.14 & $2.564\times10^{12}$ & 1:0 & 2.40 & 2.09 \\
\ovii~K$\beta$ (18.627)  & 1s3p\,$^1$P$_1$--1s$^2\,^1$S$_0$ & 0.16 & $1.010\times10^{12}$ & 0:1 & 1.70 & 1.49  \\
\ovii~r (21.602)  & 1s2p\,$^1$P$_1$--1s$^2\,^1$S$_0$ & 0.72  & $3.430\times10^{12}$ & 0:1 & 9.04 & 7.88  \\
\ovii~i (21.804)  & 1s2p\,$^3$P$_0$--1s$^2\,^1$S$_0$ & $8.2\times10^{-5}$ & $3.830\times10^8$&  & \multicolumn{2}{c}{$\sim$0}   \\
\ovii~f (22.098)  & 1s2s\,$^3$S$_1$--1s$^2\,^1$S$_0$ & $2.0\times10^{-10}$ & $9.120\times10^2$&  & \multicolumn{2}{c}{$\sim$0}
\enddata
\tablecomments{The line properties are taken from AtomDB v3 (Foster et al. 2012). $f_{lu}$ is the transition oscillator strength, $\Gamma$ is the natural broadening damping factor, and ``$w_{\rm I}:w_{\rm D}$" represents the weight ratio between isotropic scattering and dipole scattering.}
\tablenotetext{$\dagger$}{\mbox{The} right (left) column of $\tau$ values is for the case where turbulence
is (not) taken into account.}
\label{tab:lines}
\end{deluxetable*}
\end{center}

A key factor in determining the effectiveness of the RS is the motion of
ions in the hot ISM. We assume that the motion consists of
a thermal component and an isotropic turbulent component and that their
combined velocity follows a Gaussian distribution in each dimension:
\begin{equation} \label{eq:gaussian-line}
P_v(v)\,dv=\frac{1}{\sqrt{2\pi}\sigma_v}
    \exp\left[-\frac{1}{2}\left(\frac{v}{\sigma_v}\right)^2\right]\,dv .
\end{equation}
Here, the velocity dispersion is given by
\begin{equation}
\sigma_v=\sqrt{\sigth^2+\sigtu^2},
\end{equation}
where $\sigth=(kT/\mu_a m_p)^{1/2}$ and $\sigtu$ are the thermal and turbulent
velocity dispersions, while $\mu_a$ and $m_p$ are the ion species' atomic weight and
the proton mass, respectively. Accordingly, seed photons are emitted with
the frequencies following  a distribution profile:
\begin{equation}
P_{\nu}(\nu)\,d\nu=\frac{1}{\sqrt{2\pi}\sigma_{\nu}}
    \exp\left[-\frac{1}{2}\left(\frac{\nu-\nu_0}{\sigma_{\nu}}\right)^2\right]\,d\nu ,
\label{e:gau-freq}
\end{equation}
where $\sigma_{\nu}=\nu_0(\sigma_v/c)$ is the standard deviation in
frequency $\nu$ around the center frequency $\nu_0$ of the line,
while $c$ is the speed of light.

The absorption cross-section of the photons is a result of the Lorentz profile
of the line convolved with the Doppler shift (e.g., Mihalas 1978):
\begin{equation} \label{eq:cross}
s(x) = \frac{\sqrt{\pi}e^2}{m_e c\,\Delta\nu_D} f_{lu}H(a,x),
\end{equation}
where $f_{lu}$ is the oscillation strength of
the lower to upper level transition in the consideration,
$\Delta\nu_D=\nu_0(\sqrt{2}\sigma_v/c)$ is the Doppler width,
and the 
Voigt function is defined as
\begin{equation}\label{e:voigt-1}
H(a,x)\equiv \frac{a}{\pi}\int^{\infty}_{-\infty}
  \frac{\exp(-u^2)}{(x-u)^2+a^2}\,du ,
\end{equation}
with $a=\Gamma/(4\pi\Delta\nu_D)$ the Voigt parameter
($\Gamma$ is the damping width, equal to the spontaneous emisson rate $A_{ul}$),
$x\equiv (\nu-\nu_0)/\Delta\nu_D$ the dimensionless frequency shift,
and $u$ the atomic line-of-sight velocity component in units of
$\sqrt{2}\sigma_v$
(which is also equal to the Doppler frequency shift in units of $\Delta\nu_D$).
Table~\ref{tab:lines} lists the parameters of the strongest
oxygen lines used in the
application presented in the next section.

The Voigt function has an approximate profile $\exp(-x^2)$ in the
line core and approaches the asymptotic curve $a/(\sqrt{\pi}x^2)$
in the line wing (e.g., Mihalas 1978; Dijkstra et al.\ 2006a).
For the sake of saving calculation time in this work, the Voigt function is approximated as
\begin{equation}
H(a,x) \approx \exp\left(-x^2\right)
   + \frac{a}{\pi^{1/2}x^2}\left[1-\exp\left(-x^2\right)\right],
\label{e:voigt-2}
\end{equation}
with a small modification to the approximation given in Mihalas (1978),
namely adding the artificial terms in the square brackets
to avoid the divergence at the line center.

The RS is only partially coherent
in the observer's frame because of the ion's motion (Dijkstra et al.\ 2006a).
The frequency of the scattered photon is given by

\begin{equation} \label{eq:ph_en}
x_f\approx x_i-\frac{{\bf v}_a\cdot{\bf k}_i}{\sqrt{2}\sigma_v}
       +\frac{{\bf v}_a\cdot{\bf k}_f}{\sqrt{2}\sigma_v}
       +g\,({\bf k}_i\cdot{\bf k}_f-1),
\label{e:energy}
\end{equation}
where
${\bf k}$ is the unit vector of the photon's propagating direction,
the subscripts ``$i$" and ``$f$" denote the quantities of
the incident and scattered photons, respectively,
${\bf v}_a$ is the velocity of the atom (ion),
and
$g=h\Delta\nu_D/(2\mu_a m_p\,\sigma_v^2)$
is the average fraction of energy transferred per scattering
to the ion due to its recoil.
For the O\,VII resonant line, $g=8\E{-5}(\sigma_v/100\km\ps)^{-1}$,
which is negligible by orders of magnitude.

\subsection{Distribution model of the Hot ISM}

Our simulation scheme is meant to be applicable to any
spatial distribution of the hot ISM in terms of its thermal, chemical,
and kinematic properties. In principle, the distribution could even
be derived from a galaxy simulation. But in the present work,
we adopt a simple model of the medium with certain symmetry and
uniformity to characterize the effects of the RS.
For such a model,
we may define an optical depth at the line center of each
transition,
$\tau=\int n_{\rm ion}(Z)\,s(0)\,dr$,
to characterize the systemic radiative transfer condition of the gas.
The integration here is from the center of the galaxy
to the infinity or to a sufficient large
off-center distance $\rcut$, beyond which the optical depth contribution is negligible.
Ions responsible for the RS are assumed to be populated in the ground states.
The ion number density of an element $Z$, $n_{\rm ion}(Z)$, is related to
the hydrogen density as
$n_{\rm ion}(Z)=\zeta(Z)\,A(Z)\,f_{\rm ion}(T)\,\nH$,
where $A(Z)$ is the solar photosphere elemental abundance with respect to
hydrogen (H), $\zeta(Z)$ is the hot ISM elemental abundance relative to solar,
and $f_{\rm ion}(T)$ is the ionic fraction of the element as a
function of the gas temperature.
For oxygen, $A({\rm O})=8.51\E{-4}$ (Anders \& Grevesse 1989).
In this work, we calculate $f_{\rm ion}(T)$ via the APEC model, assuming the CIE
for the hot ISM.

The key parameter that determines the effectiveness of the RS is the
line-center optical depth $\tau$.
In terms of determining the energy (i.e., $x$) distribution of the
line photons emerging from the medium, $\tau$
encapsulates the dependence of the RS on
the density, size of the considered region, the temperature, the elemental
abundance, and the ionic fraction of the hot ISM, as well as its
turbulent velocity. In particular, $\tau$ is inversely proportional to
the velocity dispersion $\sigma_v$. The additional dependence of the distribution on $a$ in
the second term of Eq.~(\ref{e:voigt-2}) and on $\sigma_v$ in
 the last term of Eq.~(\ref{e:energy})
are orders of magnitude weaker\footnote{
For example, $a=1.2\E{-2}$ for the 21.602\AA\ O\,VII resonant line at $kT=0.2\keV$.}
 than the  other right-hand side terms of the equations
and can generally be neglected.
The decisive dependence of the RS effect on $\tau$ will be shown below.

The spatially resolved effect of the RS bears on the specific distribution
of the hot ISM.
In the present work, we will use
the isothermal $\beta$-model (see Cavaliere \& Fusco-Fermiano
1978; Arnaud 2009):
\begin{equation}
\nH=\no\left[1+\left(\frac{r}{\rc}\right)^2\right]^{-3\beta/2},
\label{e:beta-model}
\end{equation}
where $\no$ is the gas density at the spherical center and $\rc$
the core radius. With this model, the total radial optical depth is the natural
choice
\begin{equation}
\tau(x)\equiv \int_0^{\infty}n_{\rm ion}(Z)s(x)\,dr.
\end{equation}
It can be transformed to
\begin{equation}
\tau(x)=\tau_{\rm c}H(a,x)\int_0^{\infty}\left(1+\tilde{r}^2\right)^{-3\beta/2}d\tilde{r},
\label{eq:beta-tau}
\end{equation}
where $\tau_{\rm c}\equiv n_{\rm ion,0}(Z)\,s(0)\,\rc$,
with $n_{\rm ion,0}(Z)$ denoting the ion density at $r=0$,
$\tilde{r}\equiv r/\rc$,
and the integral has an analytic solution
$\sqrt{\pi}[\Gamma(3\beta/2-1/2)/2\Gamma(3\beta/2)]$,
with $\Gamma$ the gamma function,
for $\beta>1/3$ (Ge et al.\ 2016).

\subsection{MC simulations}

In our MC simulations, seed photons are generated with random
directional angles and with the location probability according to
the volume emissivity distribution of the hot ISM.
For the isothermal and chemically uniform medium of a
spherically symmetric distribution assumed in the present work,
the positional angle of a seed photon is random, while its radius $r$
is found from the probability distribution function
\begin{equation}
R_1=\frac{\int_0^r n_e n_{\rm ion}(Z)r'^2\,dr'}{\int_0^{\rcut}n_e n_{\rm ion}(Z) r'^2\,dr'},
\label{e:R1}
\end{equation}
where $R_1\in [0,1]$ is a randomly generated number, and
$n_e$ is  the electron density. The frequency of this photon is
generated according to Eq.~(\ref{e:gau-freq}). For the $\beta$-model
distribution,  Eq.~(\ref{e:R1}) can be simplified as
\begin{equation}
R_1=\frac{\int_0^{r/\rc}G(y)\,dy}{\int_0^{\rcut/\rc}G(y)\,dy}\,,
\hspace{5mm} G(y)=\frac{y^2}{(1+y^2)^{3\beta}}.
\end{equation}

An advantage of the MC simulation is that one can ``trace" a resonant line photon at each step in their random traveling.
In the simulation we use an optical depth
step $\Delta\tau(x)/\tau_{\rm c}$,
where $\Delta\tau(x)\equiv n_{\rm ion}(Z)\,s(x)\Delta l$, 
to determine the adaptive spatial length step $\Delta l$ along
a given random direction of photon propagation. With a sufficiently small $\Delta\tau(x)$,
the hot ISM can be assumed to be uniform in each step.
For the gas with a similarity distribution like the $\beta$-model (Eq.~\ref{e:beta-model}),
the spatial stepsize can be given in a dimensionless form:
\begin{equation}
\frac{\Delta l}{\rc}=\left[1+\left(\frac{r}{\rc}\right)^2\right]^{3\beta/2}
  \frac{\Delta\tau(x)}{\tau_{\rm c}}\,H^{-1}(a,x).
\label{eq:step}
\end{equation}
The spatial stepsize and the pathlength of the photon before each scattering (which can be a sum of small steps) are thus dominated by $\tau_{\rm c}$, which is actually controlled by the total radial optical depth $\tau(x)$ (or the line-center $\tau$ for $x=0$) through Eq.(\ref{eq:beta-tau}).

After propagating each step, the photon is checked to see whether a scattering happens.
If the scattering probability $1-\exp(-\Delta\tau(x))$ is smaller than
a randomly generated number $R_2\in[0,1]$, then no RS is declared and
the photon moves a step forward along the present direction, which may continue
until it escapes or moves outside the boundary of the system. Otherwise,
a RS event is flagged, in which a photon is re-emitted in a new direction.

Each RS event changes not only the energy of the photon (Eq.~\ref{e:energy}) but also its direction. These changes
depend on the velocity of the scattering ion, which consists of three
components: the one parallel to the direction of the incident photon,
${\bf k}_i$, and the two mutually orthogonal ones perpendicular to
${\bf k}_i$. Following the method described in Zheng \& Miralda-Escud{\'e}
(2002, also see the Appendix therein), we randomly generate the parallel
component following the distribution
\begin{equation}\label{eq:parallel}
f(u_{\|})=\frac{a}{\pi}\frac{e^{-u_{\|}^2}}{(x_i-u_{\|})^2+a^2}H^{-1}(a,x_i),
\end{equation}
where $u_{\|}$ is in unit of $\sqrt{2}\sigma_v$.
This distribution accounts for the
RS probability as a function of
$u_{\|}$ and $x_i$.
Each perpendicular component is generated from a Gaussian distribution with
zero mean and standard deviation $\sigma_v$.

The redirection of the scattered photon
can be treated as a mixture of isotropic scattering (with a weight $w_{\rm I}$)
and Rayleigh (dipole) scattering (with a weight $w_{\rm D}$), and the weighted probability
distribution is given by (Hamilton 1947; Churazov et al.\ 2010)
\begin{equation}
R_3=\int^{\mu}_{-1}\left[\frac{1}{2}w_{\rm I}+\frac{3}{8}(1+\mu'^2)w_{\rm D}\right]\,d\mu',
\end{equation}
where $R_3$ is a random number $\in[0,1]$ and $\mu$ is the cosine
of the angle between the incident and outgoing directions of the photon.
The weights $w_{\rm I}$ and $w_{\rm D}$ for the considered oxygen lines are given in Table~\ref{tab:lines}.
The relevant quantities are treated by coordinate rotation and
Lorentz transformation between the observer's frame and
the atom's rest frame in which the incident direction is the symmetric axis.

We record all the positions,
directions, and energy information of simulated photons from their
initial emission through all RS events before
escaping from the system. This complete record allows for the
post-processing of the simulation, even with the inclusion of
additional processes such as photo-electric absorption by cool gas, and
for generating mock observations, as will be demonstrated in the
follow-up paper (Paper II). 

For the resonant line emission of the hot gas,
the terms with $a$ and $g$ are negligible in the Voigt function
(\ref{e:voigt-2}) (hence also in the dimensionless positions of the photon
(Eq.~\ref{eq:step})) and the photon frequency change (Eq.~\ref{e:energy}), respectively.
Therefore, the MC simulation in terms of dimensionless frequencies and dimensionless lengths,
with the characteristic $\tau$ given, is essentially {\it irrespective of specific lines and the gas velocity dispersion}.

\section{Test Simulations and Results}\label{sec:test}

\begin{figure}[tbh!]
\centerline{
\includegraphics[scale=0.8]{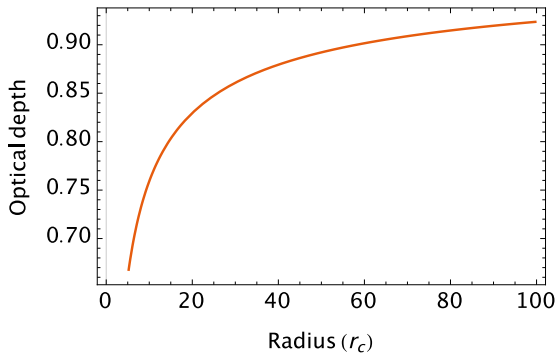}}
\caption{
Scaled radial line-center optical depth $\tau$ (in units of the maximum value at $\infty$) of the adopted  $\beta$-model hot gas ($\beta$= 0.5) as a function of the dimensionless radius over which the integration is made from the center of M31 (see \S\ref{sec:test} for details).
}
\label{f:tau-r}
\end{figure}

We run a set of simulations for a spherical $\beta$ model distribution of gas to test our scheme,
compare the calculation with the existing iterative algorithm,
and make new application to the RGS data of the M31 bulge.
The diffuse soft X-ray emission from the bulge is well characterized
with model parameters $\beta=0.5$,
$\rc=54''$ (or 0.2\,kpc at a distance 780\,kpc), and $\no\sim0.1\cm^{-3}$
(Liu \etal\ 2010; Li \etal\ 2009).
We adopt an outer
boundary at $\rcut=100\rc$, which is typically far beyond
region where the gas distribution can be modeled or the spectral data can be
extracted (\S\ref{s:app}).
The outer region ($r>\rcut$) would contribute only
a small fraction (less than a few percent) of the total optical depth
integrated to infinity (see Fig.~\ref{f:tau-r}).
Considering the drastical change of gas density at various
radii in the $\beta$-model, an upper cutoff $\Delta l_{\rm max}/\rc=0.01$
is set for the stepsize while $\Delta\tau(x)/\tau_{\rm c}=0.001$ is used.

We first carried out a test simulation
assuming a $\beta$-model of gas with $\beta=2/3$.
This simulation is used for the comparison with
the calculation under the iterative method introduced by Gilfanov (1987)
(see Appendix~\ref{appendix}).
For the comparison of the two methods, some treatments in our simulations
are simplified
(instead of those described in \S\ref{sec:modeling}).
As shown in Appendix~\ref{appendix}, the radial surface profiles
and the line profiles produced with the
MC simulation are consistent with those produced with the iterative method
for representative $\tau$ values.
We note that iterative results from Gilfanov et al.\ (1987) have been compared to the results of MC simulations in previous works (e.g., Sazonov et al.\ 2002 and consequent papers); our test is in essence also consistent with
the previous MC results.

We then run simulations with the considerations described in \S\ref{sec:modeling}
for a $\beta$-model of the hot plasma
distribution suitable for the M31 bulge.
Parameters $a$ and $g$ for the OVII resonant line at $kT=0.2$\,keV are used,
although the last terms in both Eq.(\ref{e:voigt-2}) and Eq.(\ref{eq:ph_en}) are essentially negligible.
For each simulation, if not specially stated, $10^6$ photons are run as a compromise
between a reasonably good statistical quality and a moderate computational time.

\begin{figure*}[tbh!]
\centerline{ 
\includegraphics[scale=0.6]{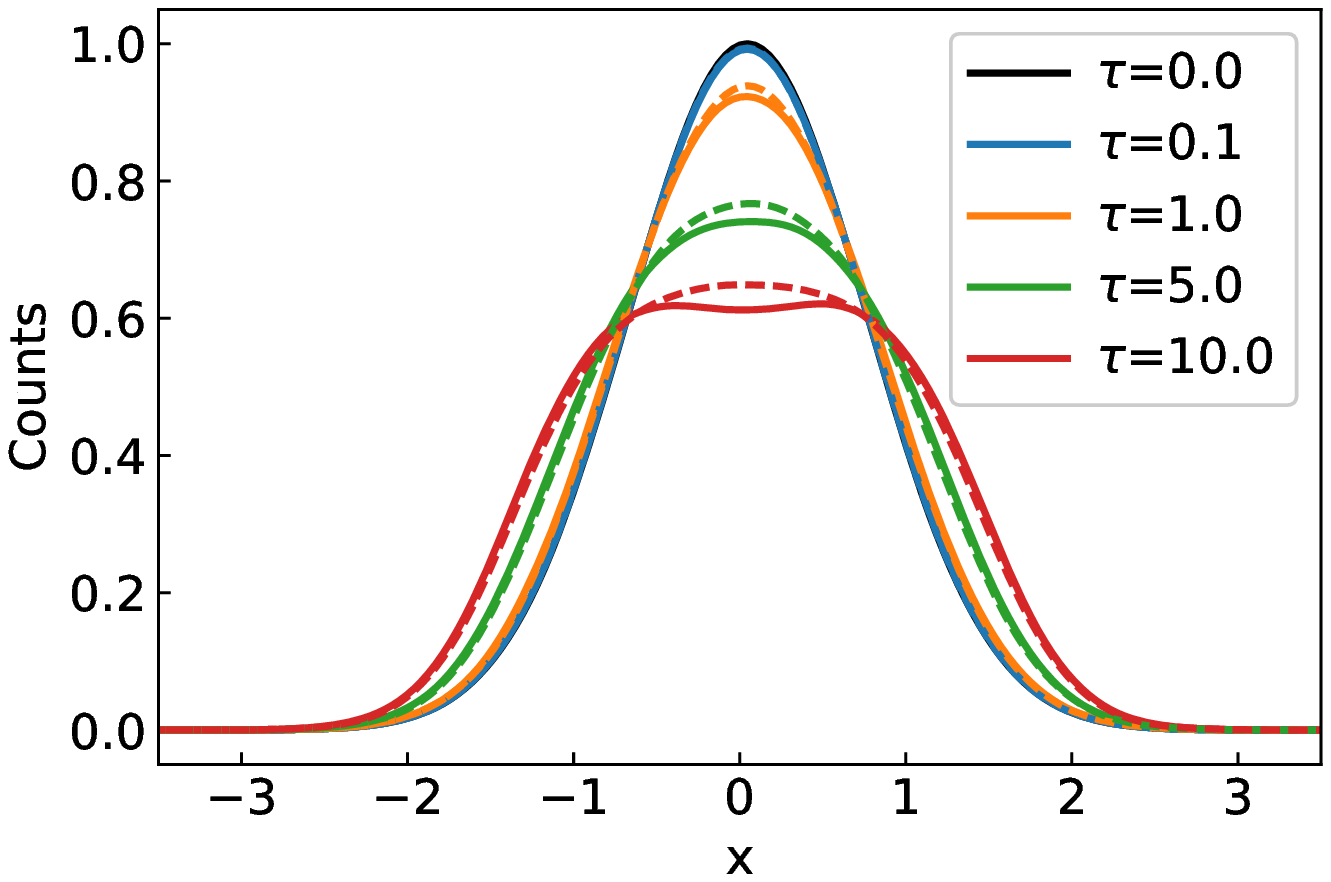}
\includegraphics[scale=0.6]{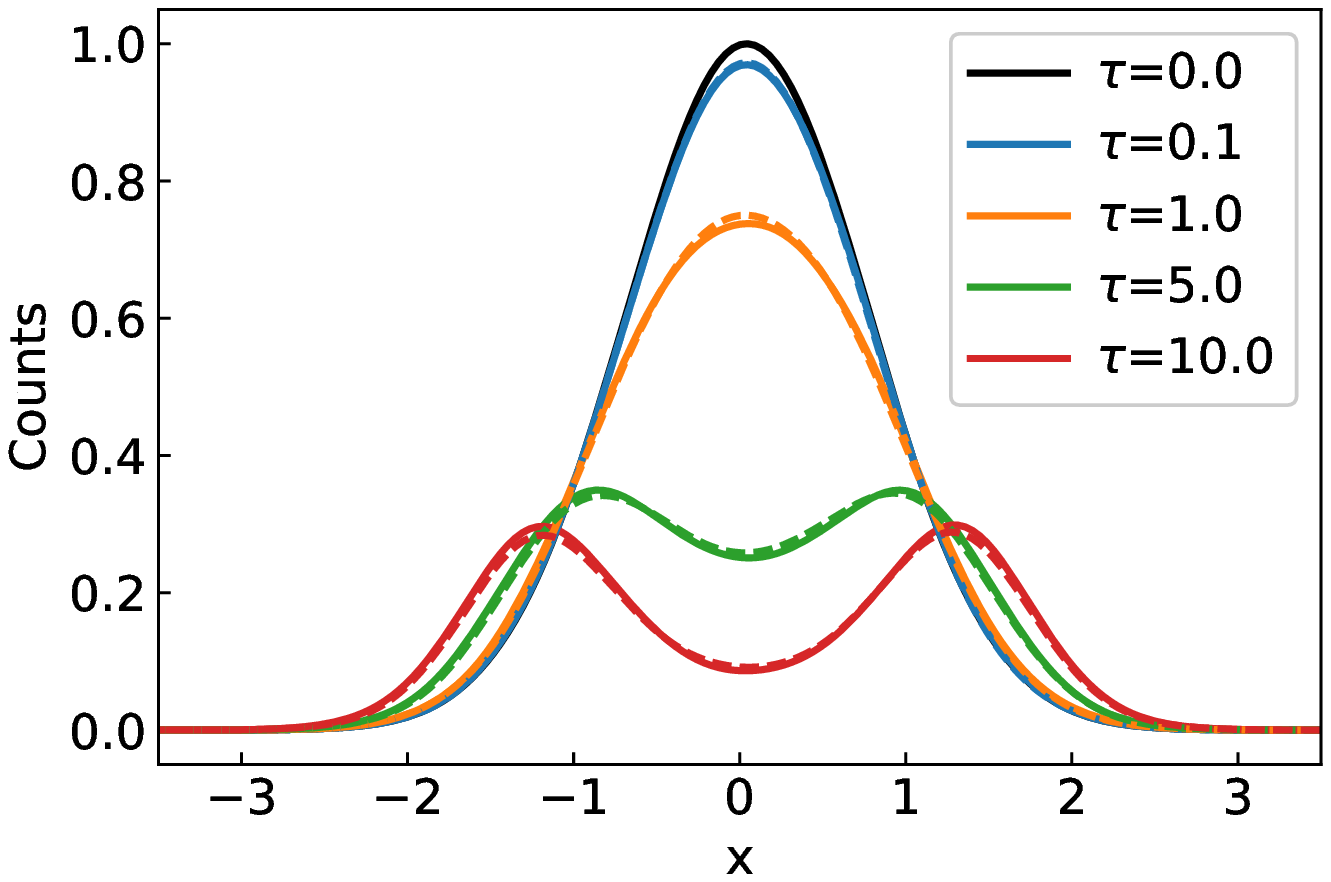}}
\caption{
{\it Left}: Line profiles of resonant photons from the $\beta$-model gas
with $\beta=0.5$ and $\rcut=100\rc$
for various $\tau$ values.
The dashed and solid lines are for isotropic and dipole scatterings (in the atom's rest frame), respectively.
{\it Right}: Same as in the left panel except for photons emerging
only from within the inner region of a projected radius $5\rc$.
A 3-pixel ($x$ interval of 0.13) Gaussian smoothing is applied to reduce the noise of the simulated data.
}
\label{f:line_profile_in_x}
\end{figure*}

Fig.~\ref{f:line_profile_in_x} presents line profiles of the photons emerging from
the RS for various representative optical depth values from the entire galaxy
and from within the projected central $5\rc$ radius, respectively.
Both isotropic and dipole scattering cases are shown.
For the same $\tau$, the line profiles differs very slightly between the two cases.
In each case, the line profile, as an explicit function of $x$ (other than of $\nu$),
changes only with $\tau$.
With the increase of $\tau$, the
core of the line gets increasingly reduced while the wing widens.
Such effect is more evident in the inner region (right panel of
Fig.\ref{f:line_profile_in_x}),
where double-peak pattern appears when $\tau \ga5$.

\begin{figure}[tbh!]
\centerline{
\includegraphics[scale=0.6]{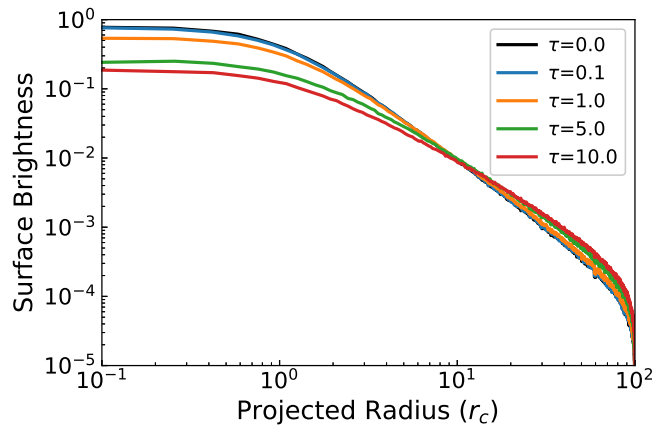}
}
\caption{
Projected radial surface brightness profiles of the hot gas line emission predicted by the adopted $\beta$-model with $\beta=0.5$ and $\rcut=100\rc$ for various $\tau$ values.
}
\label{f:rp_tau}
\end{figure}

Fig.~\ref{f:rp_tau} represents the radial surface brightness profiles
of the resonant line for various $\tau$ values.
The brightness profiles are shown for both isotropic and dipole
scatterings: there is almost no difference between the two cases, however.
With the increase of $\tau$, 
the brightness in the central projected region within 5--10$\rc$ (1--2\,kpc) drops,
while the emission from the outer region is enhanced.
Both distortions are the results of the RS-resultant random walk of photons in space.

\begin{figure}[tbh!]
\centerline{
\includegraphics[scale=0.6]{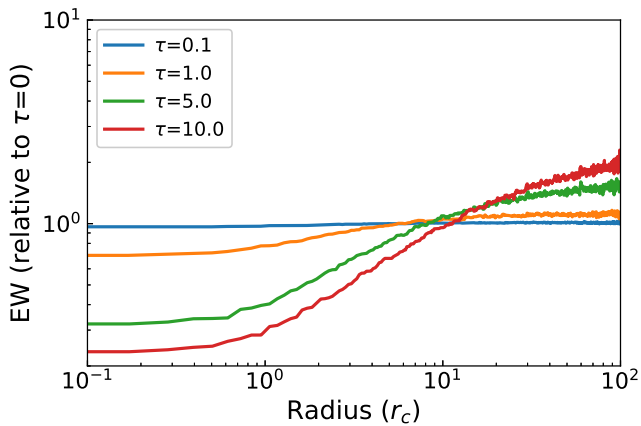}}
\caption{
Projected radial profiles of the equivalent width of the line emission predicted by
the $\beta$-model with $\beta=0.5$  and $\rcut=100\rc$, scaled with the profile of $\tau=0$.
}
\label{f:prof_ew}
\end{figure}

Fig.~\ref{f:prof_ew} gives the radial profiles of the equivalent width (EW)
of the resonant line emission relative to the ones for $\tau=0$.
With increasing $\tau$, the EW diminishes in the central
projected region and rises in the outskirt, which reflects the behavior of
the line emission brightness.

\begin{figure}[tbh!]
\center{\includegraphics[width=1\linewidth]{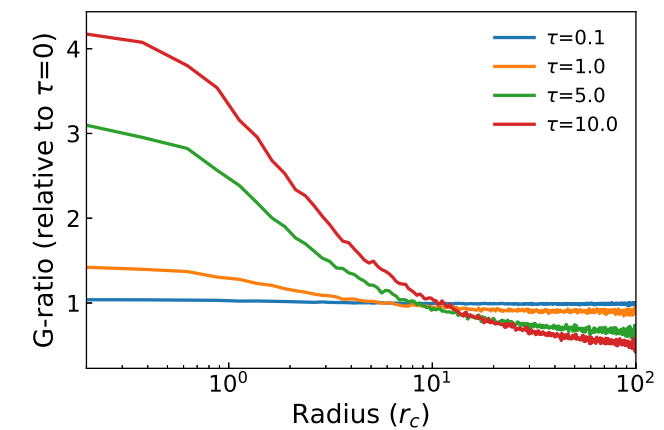}}
\caption{
Projected radial profiles of the K$\alpha$ triplet G-ratio of helium-like ions as predicted by the $\beta$-model with $\beta=0.5$ and $\rcut=100\rc$, scaled with the profile of $\tau=0$.
}
\label{f:G-ratio}
\end{figure}

The spatial redistribution of resonant line photons accordingly changes
the specific intensity of the line and hence the surface distribution of
the diagnostic G-ratio for the so-called K$\alpha$ (or He$\alpha$) triplet of
a helium-like ion, such as OVII.
The G-ratio is defined as the ratio of the sum of intercombination and forbidden components to the
resonant component (Gabriel \& Jordan 1969).
Fig.~\ref{f:G-ratio} shows the radial profiles of the G-ratio
for the $\beta$-model gas for various $\tau$ values, which are relative to the ones for $\tau=0$.
Due to the RS effect, G-ratio can be considerably elevated (e.g.,
a few times)
in the central projected region and diminished
(even by half)
in the outskirt.
Ideally, the G-ratio of the OVII triplet is close to 1 for a hot CIE gas at $\sim2\E{6}$\,K
which typifies the temperature of the M31 bulge (see \S\ref{s:app}).
Therefore, the RS effect may raise the OVII G-ratio to a value substantially higher than 1
within a few $\rc$ and suppress it to below 0.5 near the boundary.

\section{Application to the RGS spectrum of the M31 bulge}\label{s:app}

\subsection{RGS Data Reduction}

Based on the above results (e.g., Fig.~\ref{f:rp_tau}), we further build a spectral model to account for the RS effect, allowing for a direct fit to an RGS spectrum.
The RGS is a slit-less spectrometer and is sensitive to photons in the 0.3-2 keV range.
For a point-like source, the spectral resolution of the RGS is $\lambda/\delta\lambda \sim 400$ at $\sim 20$ \AA\ (for details, please refer to the \XMM\ Users' Handbook).
For an extended source, one may directly obtain useful 1-D spatial information over a width of $5'$ in the cross-dispersion direction.
In the dispersion direction, the RGS covers the entire field of view of the telescope mirrors, although the effective area decreases significantly at large off-axis angles.
A spatial displacement along the dispersion direction translates to an apparent (1st-order) wavelength shift of 0.138 \AA~per arcmin.
Therefore, for an emission region with a moderate extent (e.g., $\lesssim 2'$), its integrated RGS spectrum still has a spectral resolution substantially better than that of X-ray CCD data.

\begin{deluxetable}{lccccc}
\tablecolumns{6}
\small
\tablewidth{0pt}
\tablecaption{36 XMM-Newton/RGS observations on M31's center (2000-2012)}
\tablehead{\colhead{ID} & \colhead{RA} & \colhead{Dec} &  \colhead{P.A.} & \colhead{$t_{Exp}$} & \colhead{$t_{eff}$} }
\startdata
0109270101 & 10.680324 & 41.266006 & 76.1 & 57.9 & 33.5  \\
0109270501 & 10.680122 & 41.266124 & 76.1 & 10.6 &4.7  \\
0112570101 & 10.679958 & 41.259701 & 249.9 & 64.3 & 62.8  \\
0112570401 & 10.679071 & 41.266209 & 78.3 & 46.0 & 33.9  \\
0112570601 & 10.680028 & 41.260037 & 257.0 & 13.3 & 13.0  \\
0112570701 & 10.680006 & 41.259403 & 249.9 & 4.5 & 3.3  \\
0405320501 & 10.685502 & 41.272990 & 71.7 & 21.9 & 20.3  \\
0405320601 & 10.684280 & 41.273203 & 51.3 & 21.9 & 19.1  \\
0405320701 & 10.686068 & 41.265342 & 252.4 & 15.9 & 15.9  \\
0405320801 & 10.685275 & 41.265725 & 243.2 & 13.9 & 13.9  \\
0405320901 & 10.685723 & 41.266047 & 231.8 & 16.9 & 16.9  \\
0505720201 & 10.685294 & 41.265129 & 253.8 & 27.5 & 27.5  \\
0505720301 & 10.684693 & 41.266217 & 247.9 & 27.2 & 27.1  \\
0505720401 & 10.685944 & 41.265534 & 242.2 & 22.8 & 22.6  \\
0505720501 & 10.685340 & 41.266297 & 236.6 & 21.8 & 20.5  \\
0505720601 & 10.686532 & 41.265788 & 230.8 & 21.9 & 21.9  \\
0551690201 & 10.685436 & 41.266376 & 252.7 & 21.9 & 21.8  \\
0551690301 & 10.684949 & 41.265549 & 246.9 & 21.9 & 21.7  \\
0551690401 & 10.685264 & 41.266371 & 243.5 & 27.1 & 9.4  \\
0551690501 & 10.686075 & 41.266356 & 236.8 & 21.9 & 21.3  \\
0551690601 & 10.685709 & 41.265679 & 232.1 & 26.9 & 19.2  \\
0600660201 & 10.684718 & 41.265592 & 254.1 & 18.8 & 18.7  \\
0600660301 & 10.685422 & 41.265350 & 248.2 & 17.3 & 17.3  \\
0600660401 & 10.685428 & 41.266005 & 243.7 & 17.2 & 17.2  \\
0600660501 & 10.685658 & 41.266166 & 238.1 & 19.7 & 19.5  \\
0600660601 & 10.686281 & 41.265839 & 233.5 & 17.3 & 17.3  \\
0650560201 & 10.684482 & 41.265614 & 255.5 & 26.9 & 26.9  \\
0650560301 & 10.685331 & 41.265653 & 249.5 & 33.4 & 33.3  \\
0650560401 & 10.685800 & 41.265618 & 243.8 & 24.3 & 22.1  \\
0650560501 & 10.685515 & 41.266252 & 238.3 & 23.9 & 23.9  \\
0650560601 & 10.685947 & 41.265886 & 232.5 & 23.9 & 23.8  \\
0674210201 & 10.684554 & 41.266374 & 254.5 & 20.9 & 20.8  \\
0674210301 & 10.684424 & 41.266428 & 248.5 & 17.3 & 17.3  \\
0674210401 & 10.684782 & 41.266029 & 244.0 &19.9  & 19.9  \\
0674210501 & 10.685569 & 41.265826 & 240.7 & 17.3 & 17.3  \\
0674210601 & 10.684913 & 41.265902 & 235.0 & 26.0 & 20.4  \\
\enddata
\tablenotetext{*}{Columns are: Observation ID, Right Ascension (degree), Declination (degree), Position Angle (degree), Exposure Time (ks), and Effective Time (ks).}
\label{tab:log}
\end{deluxetable}

\begin{figure}[htbp] 
 \centering
       \includegraphics[angle=0,width=0.9\linewidth]{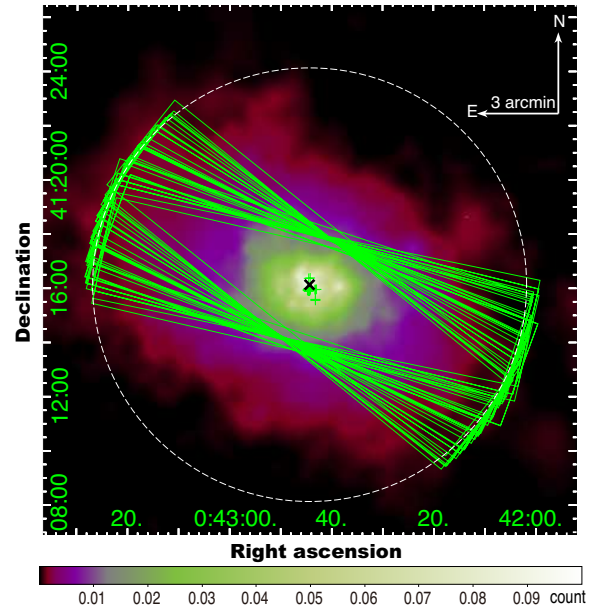}
 \caption{Chandra/ACIS 0.5-2 keV image of the diffuse emission intensity in the M31 bulge region. The white dash circle with a radius of $8'$ ($\sim$1.8 kpc) outlines a region containing about 95\% of the emission. The green rectangles ($16'\times4'$) represent the dispersion directions and the widths of our adopted spectral extraction regions of the 36 RGS observations and the green crosses show their telescope pointing centers. The black cross marks the position of the central supermassive black hole of the galaxy.}
 \label{fig:bulge}
\end{figure}

Table~\ref{tab:log} lists 36 \XMM/RGS observations toward the bulge region of M31, whose dispersion directions and the widths of extracting regions ($\sim4'$, or 920 pc) are shown in Fig.~\ref{fig:bulge}.
The Science Analysis System (SAS, v14.0) and current calibration files
are used for the data reduction.
After removing the intense background-flare periods,
the remaining total effective exposure time is 766 ks.
We use the standard `{\tt rgsproc}' script to extract the RGS spectra of individual
observations
of M31
(R.A.$=00^{\rm h}42^{\rm m}44^{\rm s}.237$ and
Dec.$=+41^{\circ}16'11''.63$, J2000) as a common dispersion reference position.
The corresponding background spectra are generated from blank-sky
events. These spectra from individual observations are combined with
the `{\tt rgscombine}' script.

\subsection{Spectral analysis}

To study the diffuse hot gas, we also need to estimate the contribution from two unrelated components in the bulge: (1) bright point sources, mostly low-mass X-ray binaries, and (2) weak unresolved point sources, mainly cataclysmic variables (CVs) and coronal active binaries (ABs).
Empirically, we model the bright-point-source component with a power law.
The X-ray emission of the CVs and ABs arises primarily from optically thin thermal plasma with temperature similar to that of the diffuse hot gas.
So the separation of their contribution is not straightforward spectroscopically.
Instead, we characterize the contribution with two {\sl fixed} APEC models of the characteristic temperatures of 4.6 keV and 0.38 keV, respectively.
These values are obtained from the spectral fitting to the unresolved emission from the small dwarf galaxy M32, which should not contain any significant diffuse hot gas.
The luminosity of this stellar X-ray emission is approximately proportional to that
of the stellar Ks-band emission: $L_{0.5-2\,{\rm keV}}/L_{\rm Ks}=(4.7\pm0.4)\times10^{27}\,{\rm erg\,s^{-1}}\,L_{\odot}^{-1}$ (Ge et al. 2015).
Using the 2MASS\footnote{https://www.ipac.caltech.edu/2mass/index.html} Ks-band data in
the typical rectangular region ($16'\times4'$) where the RGS spectra are extracted (Figure~\ref{fig:bulge}), the X-ray luminosity, which determines the normalization parameters of the APEC models, is scaled accordingly.
The spatial distribution of the contribution follows the extended stellar distribution traced by the Ks-band emission and is used to generate the broadened line profiles of the RGS spectra as described in Sec.\ 4.1.
Specifically, we convolve the angular structure function of the Ks-band image in the rectangular region with the two APEC models via the XSpec script {\tt rgsxsrc}, which uses the central supermassive black hole's coordinates and the position angle of 250$^{\circ}$.
As a result, the contribution of the unresolved point sources to the RGS spectrum is determined.

We start with characterizing the X-ray emission from diffuse hot gas with a single-temperature APEC model.
Unlike the unresolved-point-source component whose RGS line profile is modeled with the extended stellar distribution, the line profile of the hot gas component is determined by the spatial distribution of the gas itself.
Figure~\ref{fig:bulge} shows the 0.5--2 keV image of the diffuse emission intensity in M31, which is constructed from mosaicking 31 \chandra/ACIS observations (Li \& Wang 2007),
with the unresolved point source contribution scaled according to
the Ks-band image and subtracted.
The convolution is again carried out with {\tt rgsxsrc}.
The foreground absorption with a column density of $6.7\times10^{20} \,\rm{cm^{-2}}$ (Dickey \& Lockman 1990) is assumed for all the spectral components: the bright point sources, the CVs and ABs, and the diffuse hot gas.

\begin{figure*}[htbp] 
 \centering
        \includegraphics[angle=0,width=0.65\textwidth]{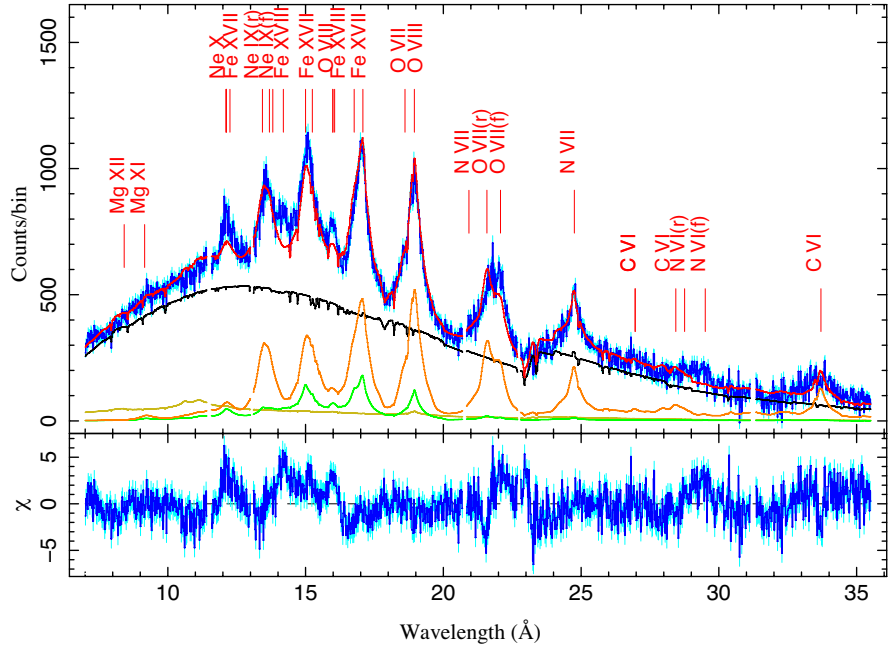}
            \caption{The combined RGS spectrum of the M31 bulge and the best-fit APEC model of optically thin plasma (red). The black curve is for the bright point sources, the green and yellow curves are for the unresolved ABs and CVs, and the orange curve is for the hot gas described with a single-temperature APEC model, respectively.}
\label{fig:apec}
\end{figure*}

\begin{figure*}[htbp] 
 \centering
        \includegraphics[angle=0,width=0.9\textwidth]{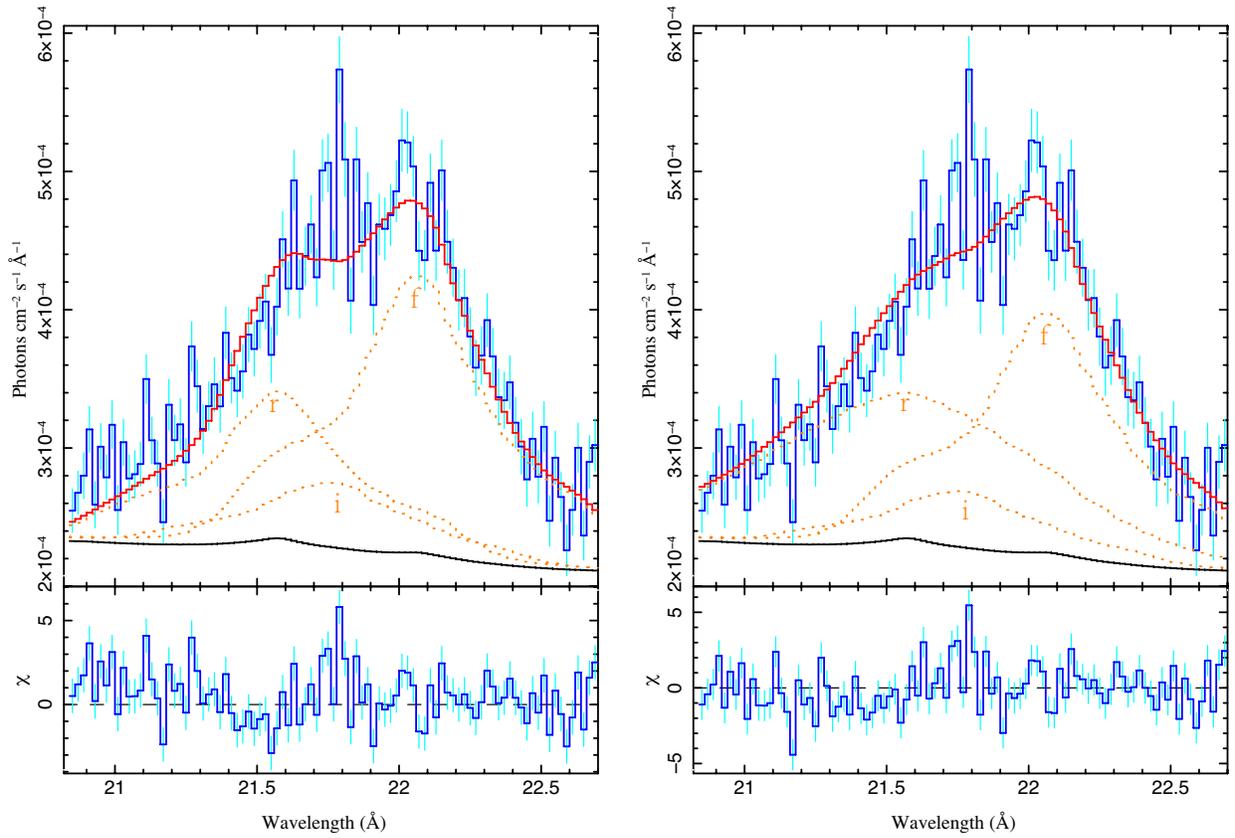}
            \caption{Model fits to the \ovii\ complex. The left panel shows the fit with fixed intrinsically narrow Gaussians convolved with the angular structure function of the hot gas distribution, while the right panel has the \ovii~r line fitted with a broader Gaussian, which leads to a decrease of the $\chi^2$ from 281 (left) to 241 (right) with 95 data bins.
            In both panels, the black curves represent the continuum, while the dotted lines labeled with ``r", ``i", and ``f" are the modeled spectra of
            the \ovii~resonant, intercombination, and forbidden lines, respectively.}
\label{fig:o7_gauss}
\end{figure*}

Figure~\ref{fig:apec} presents the best-fit APEC model
of the RGS spectrum of the M31 bulge with a temperature of $\sim$0.23 keV.
There are deviations: e.g., the \ovii~f line flux exceeds the model prediction significantly, as already noticed in Liu et al. (2010), which cannot be alleviated by simply adding more CIE plasma components.
Here, we focus on the \oviii~Ly$\alpha$ and \ovii~K$\alpha$ line complex (18--23 \AA) in the spectrum.
By dealing with transitions all from a single element, we avoid the complication that may be introduced by the uncertainties in the hot gas metal abundances.

\begin{figure*}
\centerline{
\includegraphics[width=0.5\textwidth]{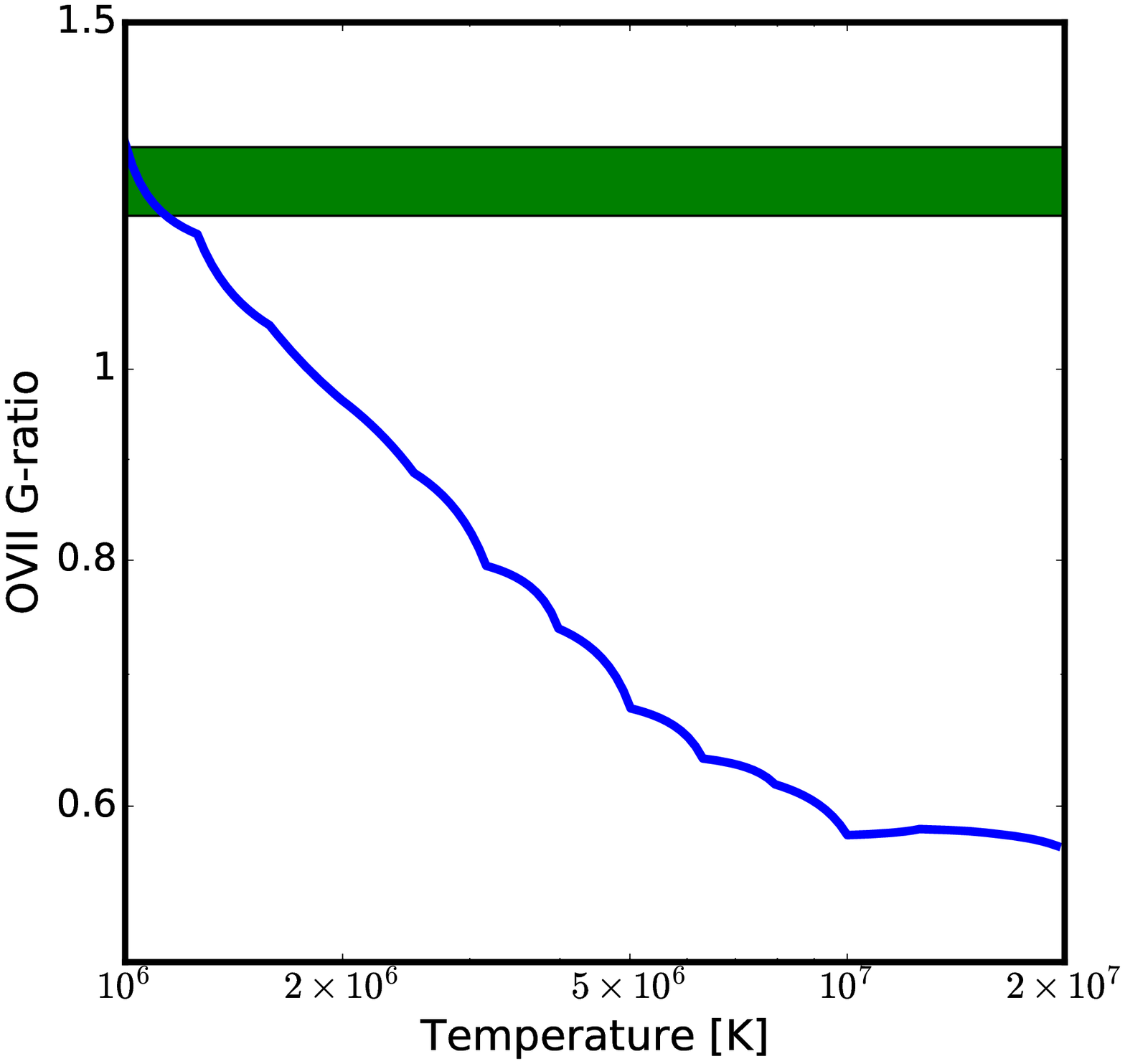}
\includegraphics[width=0.5\textwidth]{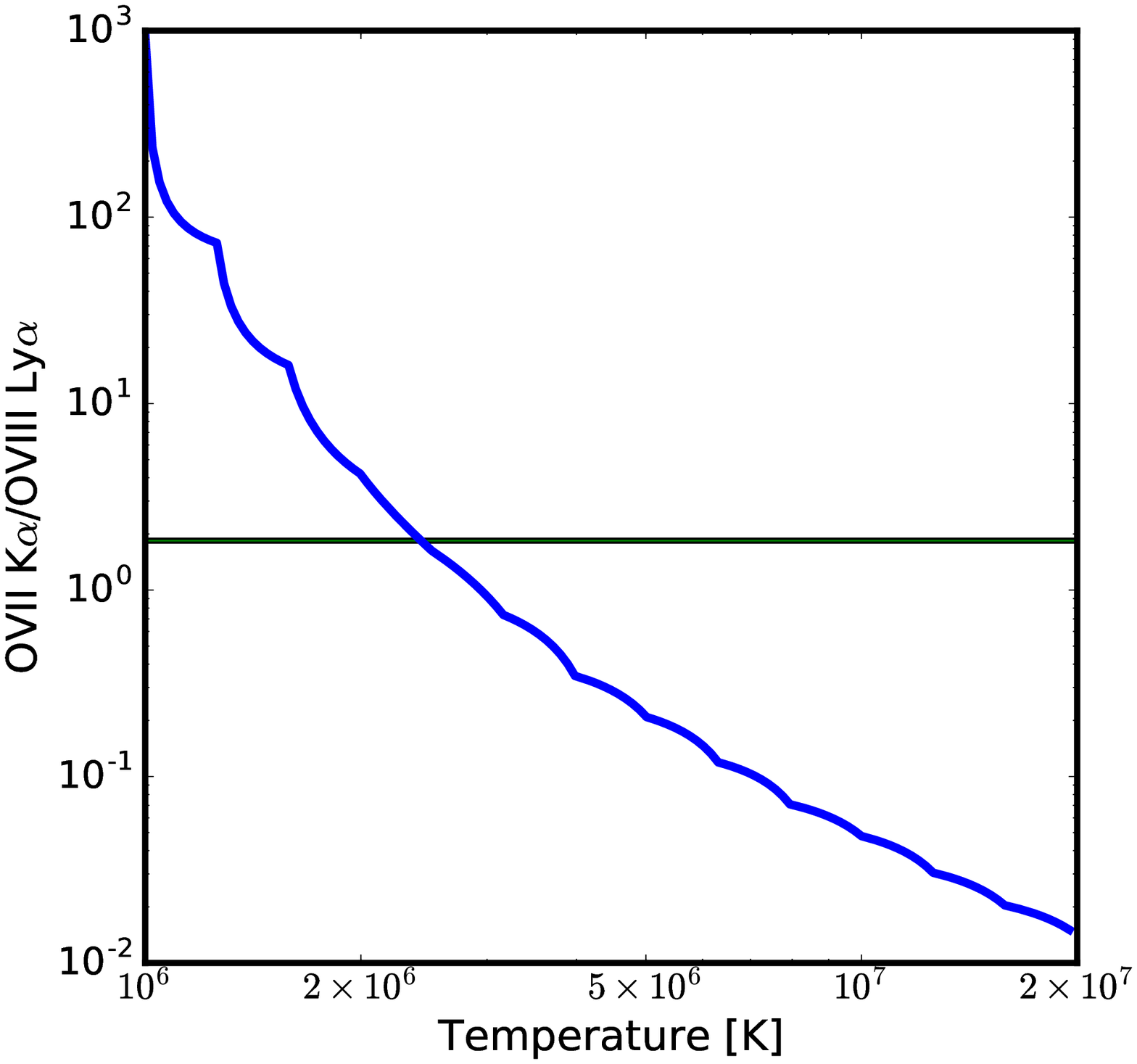}
}
\caption{
The \ovii\ K$\alpha$ triplet G-ratio ({\it left panel}) and the \ovii~K$\alpha$/\oviii~Ly$\alpha$ ratio ({\it right pabel}) as functions of the temperature of CIE plasma, where the horizontal green bars mark the measured ratios from the RGS spectrum.
}
\label{f:ratios}
\end{figure*}

To achieve a quantitative analysis, we measure the intensities of individual prominent oxygen lines (Table 1).
With the background radiation, including the emission from all point sources and the continuum emission from the hot gas, fixed to the above best-fit model,
each  line is characterized with a Gaussian, which
is convolved with the angular structure function of the diffuse X-ray image via {\tt rgsxsrc}.
The flux of the \ovii~intercombination line is scaled to that of the forbidden line with a fixed value of 0.225, which is nearly constant in the temperature range between $10^6$--$2\times10^7$ K.
Because the wavelength separation between the two \oviii~lines is less than 0.01 \AA, which cannot be distinguished in the RGS spectrum, they are jointly fitted with one single Gaussian.
The fitting results as shown in Figure~\ref{fig:o7_gauss}a give the \ovii~G-ratio as $2.39\pm0.21$, which is significantly larger than what is expected in a CIE plasma (Fig.~\ref{f:ratios}), and the \ovii~K$\alpha$/\oviii~Ly$\alpha$ ratio as $1.68\pm0.08$, which may be reproduced by a CIE plasma of a temperature $\sim2.5\times10^6$ K.
However, the blue wing of the \ovii~r line is not well fitted and appears broader than the model profile.
\begin{figure*}[htbp]
\centerline{
\includegraphics[angle=0,width=0.7\textwidth]{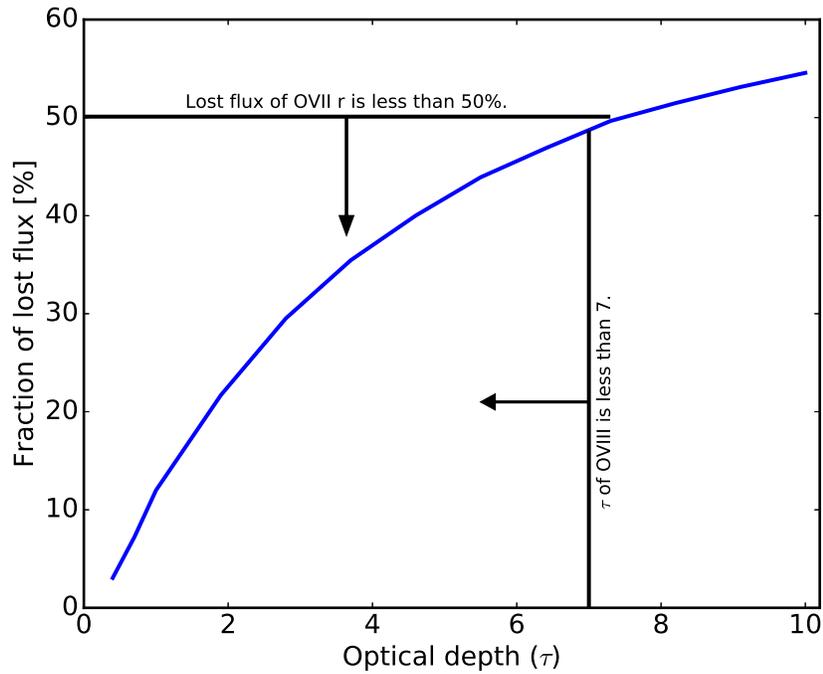}
}
\caption{
The RS flux loss in our spectral extraction region vs.\ the radial optical depth $\tau$ according to our MC simulation. We estimate that the $\tau$ value of the \oviii-L$\alpha$ line is less than 5 according to our modeled total column density and that the flux loss of the \ovii\ K$\alpha$ r line is less than 50\% from the measured G-ratio.
}
\label{f:lost}
\end{figure*}

Both the broader \ovii~r line and the high \ovii~G-ratio  may be naturally explained by the RS effect, which primarily redistributes the line photons from the inner to the outer regions.
This redistribution is opacity-dependent, and as a result, some of the OVII r photons with higher opacities can be scattered
out of the extraction region.
This leads to the decrease of the line flux (e.g., Fig.~\ref{f:rp_tau}) or the increase of the G-ratio in the region (e.g., Fig.~\ref{f:G-ratio}).
The changed angular structure function of the extended OVII r emission map in turn leads to a broader OVII r line in the RGS spectrum.
While the above analysis assumes intrinsically narrow Gaussians (with the widths fixed to 0.001\AA), we allow the width
to vary in the spectral fit, especially to the blue wing of the \ovii~r line (Figure~\ref{fig:o7_gauss}b).
The fit to the red wing at $\sim21.8$ \AA\ is also improved, although there is still an apparent excess in the RGS spectrum with a significance of about 2 $\sigma$. This excess cannot be explained by an enhanced intercombination line, because this would require a flux ratio between the forbidden and intercombination lines achievable for a plasma with density higher than $10^9\,{\rm cm^{-3}}$ (Porquet \& Dubau 2000).
Other weak lines (mainly OVII and OVI satellite lines) concentrate around 19 \AA\ and 21.6\AA, with little contribution to this feature.
Thus the residuals, if real, would most likely be due to other physical reasons (e.g., a substructure such as a cavity in the center, which is not modeled in the present work) or just be pure noise.
Therefore, the large FWHM of the model for the \ovii~r line (with a Gaussian width $0.35\pm0.06$\AA), which is about 2.5 times that of the \ovii~f line, is very likely to result from the RS effect.
With this new fit, we get the \ovii~G-ratio as $1.25\pm0.10$ and the \ovii~K$\alpha$/\oviii~Ly$\alpha$ ratio as $1.84\pm0.09$.
This latter ratio is grossly inconsistent with the CIE plasma of $\sim2.4\times10^6$\,K, which should have the \ovii~G-ratio of $\sim0.8$ (Fig.~\ref{f:ratios}).

If this RS effect is indeed important in changing the resonant line fluxes, the above temperature estimate from the line intensity ratio could then be strongly biased, however.
Our adopted $\beta$-model distribution gives a total hydrogen column density of the hot gas to be $\sim1.5 \times10^{20}$ \cmsq.
Assuming a solar abundance (Anders \& Grevesse 1989) for now, the corresponding column density of oxygen is $1.1\times10^{17}$ \cmsq.
In this case, the $\tau$ value of \oviii~Ly$\alpha$ is $\lesssim$ 7 for a CIE plasma at any temperature, even when the turbulent dispersion is negligible.
This means that at most $\sim 50\%$ of the \oviii~flux may be scattered out of the extraction region according to our MC simulation (Fig.~\ref{f:lost}).
The $\tau$ value of the \ovii~r line, on the other hand, can be $\gtrsim 10$ if the temperature is below 0.2 keV and even up to 70 around 0.1\,keV. 
However, since the \ovii~G-ratio is $\sim$1.25, roughly twice the lowest CIE value in a broad temperature range (Fig.~\ref{f:ratios}a),
the flux loss of the \ovii~r line from the extraction region should be less than $\sim$50\%.
With these flux uncertainties taken into account,
the OVII K$\alpha$/OVIII Ly$\alpha$ ratio varies between 1.2--2.6,
suggesting that the temperature of the hot gas is still in a range of 2--3$\times10^6$\,K.

Given the temperature, as well as the spatial distribution of the hot gas and the distance of M31 (780 kpc), one could also estimate the oxygen abundance from the overall intensity.
Of course, all these estimates can be strongly affected by the turbulent velocity
dispersion, which reduces the RS optical depth.
To measure the temperature, the oxygen abundance, and the velocity dispersion accurately, we need a spectrum model that accounts for the RS effect, as well as the spatial distribution of the hot gas.

\subsection{RS spectral model for the oxygen complex}

Assuming that the hot gas is isothermal, chemically uniform, and spherically-symmetric with a $\beta$-model density distribution, we separate its emission into two parts: 1) the resonant-line component that may be significantly affected by the RS, and 2) the RS-free component for the rest of emission lines and the continuum, to which the optically-thin assumption applies.
For this RS-free component (modeled like APEC minus the resonant lines), we just need to account for the spatial distribution of the plasma via a single convolution with a kernel built from the $\beta$-model.
For the RS component, we calculate the line-dependent spatial distortion and flux loss according to the changes of surface brightness (SB) profile in the MC simulation.
For computational efficiency, we pre-calculate this model in a 3-D grid of  the temperature (in the range of $10^{5.5}-10^7$~K), metal abundance (or for oxygen alone in the present simple case; 0.1--1 solar), and turbulence Mach number (0--0.4) of the plasma.
When the turbulence is isotropic, we can define the 3-D Mach number as $M=\sqrt{3}\sigtu/C_s$ where $C_{s}=\sqrt{\gamma kT/\bar{\mu} m_{\rm H}}$ is the sound speed, $\gamma=5/3$, and the mean atomic weight $\bar{\mu}$ is taken to be 0.6.

In the present application, the model includes four resonant lines: \ovii~K$\beta$, \ovii\ \ka, and \oviii\ \lya\ (consisting of two lines).
For each line and at each grid point, $\tau$ can easily be calculated, and then the convolution kernel can be derived from the simulated SB profiles and also the telescope vignetting (e.g., Fig.~\ref{f:rp_tau}).
The normalization parameter of the model is the same as that of the APEC model: i.e., $\eta=\int n_e n_H\, dV/(4\pi d^2)$ (where $d$ is the distance to M31).
Here, $\eta$ is readily estimated as $6.6\times10^{11}\,{\rm cm}^{-5}$ based on the RGS extraction region of the $\beta$-model. 
We call this table model as the RS spectral model, and all the calculations are based on the atomic database AtomDB (v3.0.2; Foster et al.\ 2012).

\subsection{Fitting and Discussion}

\begin{figure*}[htbp]
 \centering
       \includegraphics[angle=0,width=4.5in]{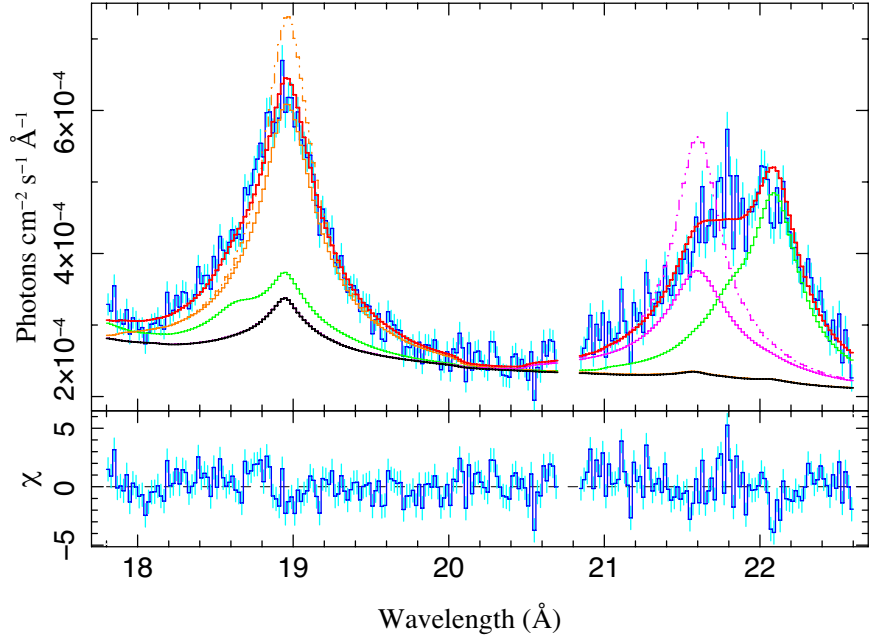}
 \caption{The RS spectral model (solid red line) best fitted to the OVIII+OVII complex of the RGS spectrum of the M31 bulge.
 Individual lines or combinations are illustrated separately: \oviii\ Ly$\alpha$ (orange), \ovii\ resonance line (purple), and \ovii\ K$\alpha$ intercombination+forbidden lines, \ovii\ K$\beta$, and other satellite lines (green). For comparison, the corresponding model resonant lines without the RS are shown (dash-dot). The black line represents the combined contribution from point sources and the hot gas continuum emission. The data at 20.75 \AA, which are uncertain because of the presence of a CCD gap, are excluded in the spectral fit here.}
 \label{fig:rgs}
\end{figure*}

Fig.~\ref{fig:rgs} presents a fit of the RS spectral model to the OVIII+OVII complex of the RGS spectrum in the 18-23 \AA~range.
Formally, the fit with $\chi^2/d.o.f =470/(233-3)\simeq2.0$ does not seem to be satisfactory.
While only statistical errors are included in the fit, systematic errors (though difficult to quantify) may be important for the spectral data, which have counting statistics,
and may well dominate in this case.
As such, we consider that the fit is quite reasonable, which constrains the temperature as $0.195\pm0.001$ keV (or $\sim2.26\times10^6$ K),
the oxygen abundance as $0.51\pm0.02$ solar value,
and the Mach number as $0.17\pm0.02$ (all at the 90\% confidence level).
With the fitted temperature, we infer the sound speed of the hot gas as $C_s \sim228$ \kmps.
The Mach number corresponds to an isotropic turbulence velocity dispersion of $40\pm4$ \kmps.
The characteristic $\tau$ values of individual resonant lines corresponding to the best-fit spectral parameters are given in Table~\ref{tab:lines}.

The above simple application of combining the simulations and spectral modeling demonstrates its feasibility. The most interesting result is the first constraint on the Mach number of the turbulence in the hot ISM of the M31 bulge.
In particular, the estimate of the Mach number depends chiefly on the relative intensities of the lines, especially the G-ratio (which is obtained from the lines of the single ion species, or the \ovii\ \ka\ triplet), and should be quite robust.
The best-fit Mach number is small, which seems to be broadly consistent with those estimated for massive elliptical galaxies (e.g., Werner et al.\ 2009; Ogorzalek et al.\ 2017) and clusters of galaxies (e.g., Hitomi collaboration 2016), within the measured uncertainties for individual objects or among them.

However, our results do depend on multiple assumptions, which need to be tested carefully.
We have assumed the $\beta$-model inferred previously.
The fitting results are generally insensitive to any small change ($\lesssim 10\%$) of the parameters in the $\beta$-model.
Parameters other than the Mach number, such as the temperature and abundance, are sensitive to various other assumptions that we have made in the modeling.
The isothermality of the hot plasma is probably most problematic.
Indeed, we find that our best-fit RS spectral model for the \ovii+\oviii\ complex cannot fit the entire RGS spectrum.
The iron emission lines in the spectrum seem to prefer a higher plasma temperature of $\sim0.6$ keV, for example.
Therefore, a more sophisticated model, phenomenological and physical, is needed to explain the spectrum and to advance our understanding of the hot gas in the M31 bulge (Zhang et al.\ 2018a, in prep).

\section{Summary}

Observational evidence is growing for the importance of the RS of soft X-ray
resonant line emission by diffuse hot plasma in nearby galaxies,
which affects the interpretation of its observed spectrum.
Taking advantage of the modern computing power, we are developing a direct
MC simulation scheme that will enable us to flexibly deal with the RS in the
plasma, in principle, with an arbitrary spatial, thermal, chemical,
and kinematic distribution. We have here reported the initial implementation of
this scheme via dimensionless calculation to an isothermal, chemically uniform, and
spherically symmetric plasma with a radial density distribution
which can be characterized by a simple $\beta$-model.
The simulation resulting from this scheme can be directly compared
with those from existing calculations.
In particular, we find that our results are consistent with those obtained from
an iterative method allowed for a simplified treatment (Gilfanov \etal\ 1987).

Our spectral model dealing with RS effects
simultaneously accounts for the optical depth-dependent spatial distortion and
strength change of the line emission, consistent with previous calculations.
This spectral model can be built for any specific spectral extraction region and
from the simulation with an assumed spatial distribution of the plasma.
We have fitted the model to the \ovii/\oviii\ complex in the RGS spectrum of
the inner bulge of M31, providing constraints on
both isotropic turbulent velocity dispersion and temperature of the plasma as $40\pm4$ \kmps\ and $0.195\pm0.001$ keV.

However, we do find that the isothermal model is too
simplistic to explain the entire RGS spectrum of the plasma. Therefore,
further refining the modeling of the
spectrum is needed to accurately measure the thermal, chemical, and kinematic
properties of hot plasma.
Extending the simulations to more physical models of the plasma distribution and to larger spectral ranges
will enable us to study the RS effect in more realistic and complicated cases
(e.g., galactic hot gaseous disks of various inclination angles).

\section{Acknowledgments}
We appreciate the anonymous referee for constructive comments
and Zheng Zheng for carefully reading the manuscript.
We thank Shikui Tang, Hui Li, Wei Sun, and Xin Zhou for various
technical helps in the study. We also thank Shawn Roberts for the help with the M31 data fit.
Y.C.\ acknowledges the supports from the 973 Program grants 2017YFA0402600 and 2015CB857100 and NSFC grants 11773014, 11633007, 11851305, 11233001, and 10725312.
S.Z.\ acknowledges the support of the China Scholarship Council and the supports from NSFC grants 11573070 and 11203080.

\appendix \section{Appendix: Comparison with Iteration Method}
\label{appendix}

Gilfanov et al.\ (1987) present an iterative method to calculate the RS effect under the assumption of small optical depth. As a general situation, the emissivity can be determined iteratively from the one with no scattering to the one with $n$th scattering:
\begin{equation}
j_{\nu}^{(n)}(r)=j_{\nu}^{(n-1)}(r)+\frac{\phi_{\nu}}{4\pi}n_i(r)\int d\Omega^{\prime}\int_{0}^{+\infty} d\nu^{\prime} s(\nu^{\prime})\int_{-\infty}^{0} d\xi\,  j_{\nu^{\prime}}^{(n-1)}(r_{\xi})e^{-\tau_{\nu^{\prime}}(\xi)},
\end{equation}
where $j^{(n)}_{\nu}(r)$ denotes the emission coefficient after
the $n$th scattering,
$\phi_{\nu}=\exp(-x^2)$ is the line profile function, $s(\nu^{\prime})$ is the cross section of the scattering, and $r_{\xi}=\sqrt{r^2+\xi^2-2r\xi\mu'}$
(with $\mu'$ the cosine of the angle between the radial vector and the incident direction).

\begin{figure*}[tbhp]
\centerline{ 
\includegraphics[scale=0.8]{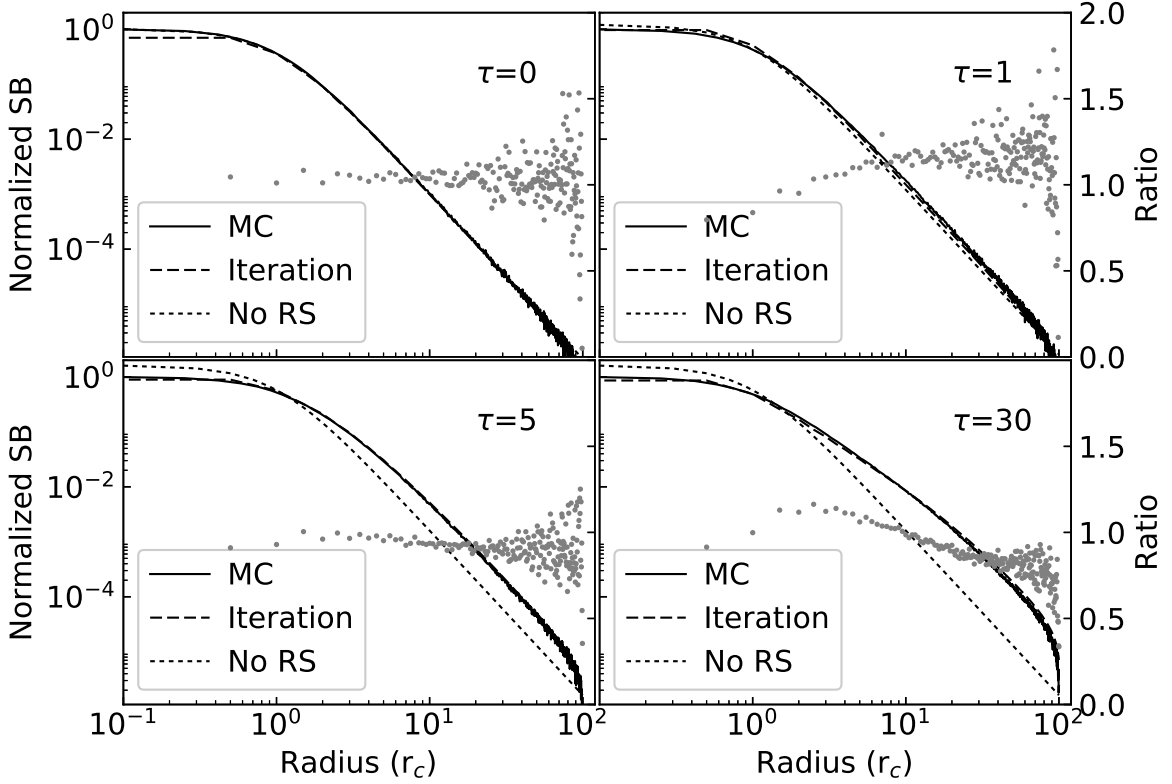}}
\caption{
Comparison of resonant line surface brightness profiles
obtained with two methods: the solid lines demonstrate the MC simulation results,
while the dashed lines demonstrate the iterative calculation results.
The grey dots represent the ratios between the MC and iteration results (with the scale shown on the right side of the figure). Both results are obtained for the hot gas distribution following the $\beta$-model with $\beta$= 2/3. The four panels show the cases of different $\tau$ values.
For reference, the brightness profile for $\tau=0$ (i.e, no RS), obtained from numerical integration of emission measure $\int n_e n_{\rm ion}(Z)\,dl$ along the line of sight, is plotted in all panels.
}
\label{fig:rp_diff}
\end{figure*}

\begin{figure*}
\centerline{ 
\includegraphics[scale=0.6]{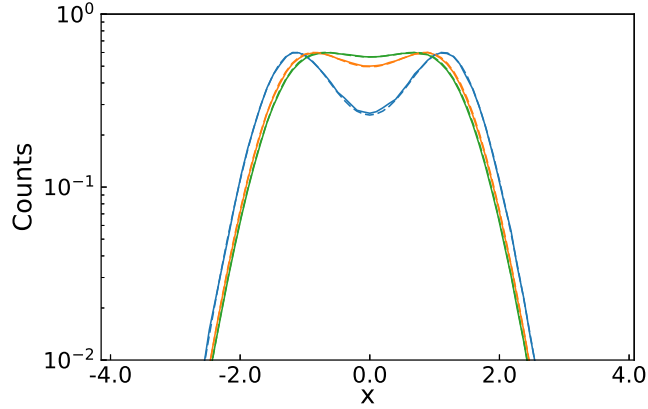}}
\caption{
Comparison of the resonant line profiles
obtained with two methods, assuming the same $\beta$-model as in Fig.\ref{fig:rp_diff}, but for $\tau= 5$ only.
The solid and dashed lines are obtained from the MC simulation and iterative calculation, respectively.
The line profiles of the emission within projected radii $1\rc$, $5\rc$, and $10\rc$ are shown in {\em blue}, {\em red}, and {\em green}, respectively.
In the simulation, $10^8$ photons are calculated. No smoothing is applied to the line profiles.
}
\label{fig:Fe_profile}
\end{figure*}

We consider the scattering of resonant line emission in an isothermal hot gas of $\beta$-model distribution with $\beta=2/3$ like that used in Gilfanov et al.\ (1987), but with a cutoff $\rcut=100\rc$.
For a comparison between the two methods, we make the MC simulation with the simplification used in the iteration calculation. Namely,
isotropic scattering is considered,
only the first term of the right hand side of Eq.(\ref{e:voigt-2}) is used for the Voigt function,
a Gaussian distribution instead of Eq.(\ref{eq:parallel}) is used for the parallel velocity component,
and a Gaussian distribution instead of Eq.(\ref{eq:ph_en}) is used for the frequency of the scattered photons.
A comparison of the radial brightness profiles obtained from the iterative method and our MC simulation method for various $\tau$ values is shown in Fig.~\ref{fig:rp_diff}.
The brightness profiles from the two methods are in a good agreement.
The ratios between the MC and iteration methods are generally close to unity,
except in the region near the boundary.
Near the boundary, the RS numbers are relatively low, which increases the statistical noise.
Moreover, the photons in our iterative calculation include those incident from outside the boundary
(leading to an increase in the intensity). In contrast, no such photon is included in the MC calculation.
The decrease in the ratios due to this difference is apparent near the boundary in the large optical depth cases, as seen in the bottom-right panel of Fig.~\ref{fig:rp_diff}.
In addition, the resolution in the iterative calculation is relatively low and is largely limited by the computing power, which also contributes to the discrepancy between the two methods.

Fig.~\ref{fig:Fe_profile} presents the line profiles for the emission within projected radii $1\rc$, $5\rc$, and $10\rc$, respectively, for the radial optical depth $\tau=5$ obtained from the two methods. The line profiles obtained from the simulation match those from iterative calculation reasonably well.
The line profile from the central $1\rc$ region exhibits
double-peak shape; similarly to Gilfanov et al.'s (1987) case for the geometric center,
the two peaks are at $x\sim\pm1.2$.
The shape of the emission line from the entire region (within $100\rc$)
is also very similar to the one obtained by Gilfanov et al.\ (1987).

\end{document}